\documentclass{elsarticle}
\usepackage[utf8]{inputenc}
\usepackage{amsmath}
\usepackage{amssymb}
\usepackage{amsfonts}
\usepackage{amsthm}
\usepackage{graphicx}
\usepackage{color}
\usepackage{hyperref}
\usepackage{tikz}
\usepackage{xspace}
\usepackage{needspace}
\usepackage{url}
\usepackage{subcaption}
\usetikzlibrary{decorations.pathmorphing}
\newtheorem{theorem}{Theorem}[section]

\newtheorem{proposition}{Proposition}
\theoremstyle{definition}

\newtheorem{algorithm}[theorem]{Algorithm}
\let\oldalgorithm\algorithm
\renewcommand{\algorithm}{\oldalgorithm\normalfont}

\begin{document}
\begin{frontmatter}
\title{Unraveling simplicity in elementary cellular automata}
\author[lif]{Pierre-Étienne Meunier\corref{cor1}\fnref{fn1}}
\ead{pierre-etienne.meunier@lif.univ-mrs.fr}
\cortext[cor1]{Corresponding author}
\address[lif]{Aix Marseille Université, CNRS, LIF UMR 7279, 13288, Marseille, France}
\date{}
\begin{abstract}
We show that a large number of elementary cellular automata are
computationally simple. This work is the first systematic classification of
elementary cellular automata based on a formal notion of computational
complexity.  Thanks to the generality of communication complexity, the
perspectives of our method include its application to other natural systems such
as neural networks and gene regulatory networks.
\end{abstract}
\end{frontmatter}
\section{Introduction}
Many computational processes can be seen as a sequence of information exchanges
between parts of the space. Moreover, a large number of natural systems, either
physical or biological, bare close similarity to algorithmic processes, in the
sense that they are dynamical processes, in which local information exchanges
play an important role, as noticed for instance by Maxwell \cite{maxwell1871},
or more recently by biologists working with DNA, or by quantum information
theorists \cite{bennett1998quantum}.
In this work, we present a method to analyze this kind of systems, and apply it
systematically to a popular class of cellular automata, showing that most of
them are computationally simple.  Our main tool to complete this task is called
communication complexity, a computational model introduced by Yao \cite{yao79}
to prove lower bounds in VLSI design (see \cite{kushilevitz97} for a full
introduction).
Cellular automata are a model of computation primarily consisting of simple
local interactions. This kind of dynamics is ubiquitous in many physical or
biological processes; developing powerful tools to analyze these objects
therefore seems an important step in the study of these systems.
Originally introduced by von Neumann \cite{neumann67} to study self-reproduction
of computationally meaningful ``organisms'', cellular automata have given rise
to a rich theory: in particular, their computational capabilities have been
extensively studied \cite{lindgren90,hanson97,kari92,moore98}. The idea of
universality \emph{modulo rescaling}, also called \emph{intrinsic universality},
has also originated in this model
\cite{surveyOllinger,Theyssier05,BoyerT09,liferokadur}, before inspiring
research in other fields, most notably in tile self-assembly
\cite{RotWin00,USA,IUSA,2HAMIU}, where it helped understand long-standing open
problems of that domain \cite{Meunier-2014}.
This notion of simulation is stronger than Turing universality, and allows
reasonings on geometric or topological properties, that are most of the time
preserved by this operation. However, it is not \emph{too} strong to be
meaningful: indeed, \emph{intrinsically universal} cellular automata have been
known for some time \cite{surveyOllinger}.
The simplest examples of cellular automata, the one-dimensional ones with two
states and two neighbors, are called \emph{elementary}. Since the simulation
works of Wolfram \cite{wolfram84}, they have received a lot of attention,
culminating in a result by Cook showing that one of these rules (called ``rule
110''), is capable of arbitrary Turing computation \cite{cook04}. His
construction was later improved by Neary and Woods \cite{woodsneary06}.
However, being able to \emph{program} a system, i.e. to use it to perform
nontrivial algorithmic operations, does not mean that we understand its
computational capabilities. On the contrary, complexity \emph{lower bounds}
require such an understanding: for example, showing that a class of logic
circuits is not able to compute some function requires a full understanding of
its capabilities, to make sure that no ``hidden trick'' allows it to compute
that function.
The idea of analyzing the communication patterns in cellular automata was first
used by D\"urr, Theyssier and Rapaport \cite{durr04}, and then applied to a
particular cellular automaton, rule 218 \cite{rap08}. More recently, a
connection to intrinsic universality was made in \cite{goles2011}.
In the present paper, we bring this method to a new level, by systematically
studying \emph{all} elementary cellular automata, proving formally, for most of
them, that their dynamics are simple, in the sense that it does not ``embed''
arbitrarily complex dynamics. More precisely, the main result of our paper is:
\begin{theorem}
  55 of the 88 non-isomorphic elementary cellular automata are not intrinsically
  universal.
\end{theorem}
\section{Definition and preliminaries}
Let $Q$ be a finite set called the set of \emph{states}.
A \emph{cellular automaton} is a map of $Q^{\mathbb{Z}}$ to itself, defined by a
\emph{local rule} $f:Q^{2r+1}\rightarrow Q$ for some integer $r$. The cellular
automaton defined by a local rule $f$ is the map $F$ such that for all $x\in
Q^{\mathbb{Z}}$, all $i\in\mathbb{Z}$,
$(F(x))_i=f(x_{i-r},x_{i-r+1},\ldots,x_{i+r})$.
An important particular case of cellular automata is the \emph{shift operator}
$\sigma$, which is the map defined for all $x\in Q^{\mathbb{Z}}$ by
$(\sigma(x))_i=x_{i+1}$ for all $i\in\mathbb{Z}$. Remark for example that any
cellular automaton commutes with $\sigma$ (and this is indeed a part of an
alternative definition, see \cite{hedlund69}).
The elements of $Q^{\mathbb{Z}}$ are usually called \emph{configurations}, and
positions inside configurations are called \emph{cells}.
Cellular automata are defined on finite configurations by extension of this
definition: the image by $F$ of a finite configuration $x$ of size $n$, is
the finite configuration $y$ of size $n-2r$, obtained by applying the local
rule of $F$ at all positions of $x$ where it is defined.
We now define a \emph{bulking} operation to compare different cellular automata:
for all integer $m>0$, let $b_m$ be the map of $Q^{\mathbb{Z}}\rightarrow
(Q^{\mathbb{Z}})^m$ defined for all $x\in Q^{\mathbb{Z}}$ and all position
$i\in\mathbb{Z}$ by $(b_m(x))_i=(x_{m i},x_{m i+1},\ldots,x_{m i+m-1})$.  For
all cellular automata $F$, all integers $t$ and $z$, the rescaling of $F$, with
parameters $m$, $t$ and $z$, is defined by:
$$F^{\langle m,t,z\rangle}=b_m\circ \sigma^z\circ F^t\circ b_m^{-1}$$
Moreover, for two cellular automata $F$ (with states $Q_F$) and $G$ (with states
$Q_G$), we say that $G$ is a \emph{subautomaton} of $F$ if there is an injection
$\phi:Q_G\rightarrow Q_F$ such that $F\circ\phi=\phi\circ G$. Intuitively, this
means that $G$ has the same behavior as $F$, but only a subset of the states of
$F$.
We say that $F$ \emph{simulates} $G$ if $G$ is a subautomaton of some rescaling
of $F$. An \emph{intrinsically universal} cellular automaton is one that simulates
any other cellular automaton.
Elementary cellular automata are those cellular automata with two states
$\{0,1\}$ and two neighbors. A classical way to enumerate them, first used
by Wolfram~\cite{wolfram84}, is to used the binary words from by the image
of their local rule, which is the integer $\sum_{0\leq i\leq 7}2^if(i)$.
Now, let $X$, $Y$ and $Z$ be three finite sets, and $f$ be any function of
$X\times Y\rightarrow Z$. The \emph{communication complexity} of $f$ is the
minimum, over $(x,y)\in X\times Y$, of the number of bits that need to be
communicated by two players, Alice and Bob, in order to compute $f(x,y)$, when
Alice knows only $x$, and Bob knows only $y$.
Formally, a \emph{communication protocol} is a binary tree, where the internal
nodes are labeled either by a function of $X\rightarrow
\{\mathrm{l},\mathrm{r}\}$, or by a function of $Y\rightarrow
\{\mathrm{l},\mathrm{r}\}$, and the leaves are labeled by a value $z\in Z$.  A
protocol $P$ \emph{computes} a function $f:X\times Y\rightarrow Z$ if for all
input $(x,y)\in X\times Y$, the leaf reached by the following procedure is labeled $f(x,y)$:
\begin{itemize}
\item Start with the current node being the root
\item If the current node $n$ is an internal node, labeled by a function
  $v:X\rightarrow\{\mathrm{l},\mathrm{r}\}$, update the current node to become
  the right child of $n$ if $v(x)=\mathrm{r}$, and its left child else.
\item If it is an internal node labeled by a function
  $v:Y\rightarrow\{\mathrm{l},\mathrm{r}\}$, update the current node to become the right
  child of $n$ if $v(y)=\mathrm{r}$, and its left child else.
\item If it is a leaf, stop.
\end{itemize}
The \emph{deterministic communication complexity} $D(f)$ of $f$ is then defined
as the depth of the least deep tree that computes $f$. By extension, for a
function $f:X^n\rightarrow Z$, for some integer $n$, the deterministic
communication complexity of $f$ is the maximal deterministic communication
complexity, over all cuts of the input: $D(f)=\max_{0\leq i<n} f_i$, where for
all $i\in\{0,1,\ldots,n-1\}$, $f_i$ is the function of $X^i\times
X^{n-i}\rightarrow Z$ defined by $f_i(x,y)=f(xy)$.
\section{Explanation of the method}
Our method is based on theorems relating communication complexity of deciding
questions on the dynamics of cellular automata, to intrinsic
universality~\cite{goles2011}. We will not state the proof of these theorems:
intuitively, they are based on the fact that communication complexity is
preserved by rescaling and the subautomaton relation, in the sense that if $F$
simulates $G$, any protocol solving a problem on $F$ can be used to solve the
same problem on $G$.
\begin{theorem}
  Let $F$ be an intrinsically universal cellular automaton. Then
  $D(\textsc{Pred}_{F,n})\in\Omega(n)$, where for all integer $n$,
  $\textsc{Pred}_{F,n}$ is the function of $Q_F^{2n+1}\rightarrow Q_F$ defined
  for all $x\in Q^{\{-n,-n+1,\ldots,n\}}$ by $\textsc{Pred}_{F,n}=(F^n(x))_0$.
\end{theorem}
The $\textsc{Pred}$ problem is probably the most natural question on cellular
automata: intuitively, it asks to predicting the evolution of the central cell
of a configuration over time. A similar problem is the following $\textsc{SInv}$
problem, asking whether finite changes in an infinite configuration remain for
arbitrarily long:
\begin{theorem}
  Let $F$ be a cellular automaton.  For all $u\in
  Q_F^{\ast}$, let $p_u\in Q^{\mathbb{Z}}$ be the infinite word defined
  for all $i\in\mathbb{Z}$ by $(p_u)_i=u_{i\mod |u|}$, and for all finite
  words $x\in Q_F^\ast$, let
  $p_u[x]$ be the infinite word equal to $x$ on $\{0,1,\ldots,|x|-1\}$,
  and to $p_u$ everywhere else.
  Then, let $\textsc{SInv}_{F,u}$ be the problem of deciding, on input $x$,
  whether there is an integer $w$ such that for all $t$, the differences of
  $F^t(p_u)$ and $F^t(p_u[x])$ are all within a part of width $w$ of
  the configuration.
  If $F$ is intrinsically universal, then there is a word $u\in Q_F^{\ast}$,
  such that $D(\textsc{SInv}_{F,u})\in\Omega(n)$,
\end{theorem}
\section{Simple cellular automata}
\begin{proposition}
  Rules 15, 51, 60, 90, 105, 108, 128, 136, 150, 160, 170 and 204 are linear,
  and thus have a protocol for $\textsc{Pred}$ in $O(1)$ (by a theorem of
  \cite{goles2011}).
\end{proposition}
\begin{proposition}
Rule 76 has a protocol in $O(1)$ for $\textsc{Pred}$.
\begin{proof}
On all configurations after one step, rule 76 behaves like rule 204,
because the only difference is on 111, which has no antecedent. Therefore,
Alice and Bob need to communicate one bit to compute the first step,
and then follow the protocol for rule 204.
\end{proof}
\end{proposition}
\begin{proposition}
Rules 0, 1, 2, 4, 8, 10, 12, 19, 24, 34, 36, 38, 42, 46, 72,
76, 108, 127, 138, 200 have a constant number of dependencies, and thus have a
protocol in $O(1)$ for $\textsc{Pred}$.
\begin{proof}
  We treat these cases independently. An argument that we will frequently use,
  is that computing the configuration after one step requires at most $2r$ bits
  of communication: Alice and Bob communicate their $r$ bits around the
  separation between their respective inputs to each other, and then compute one
  step of the local rule separately on their inputs, concatenated with the
  received bits.
\begin{itemize}
\item Rule 0 is nilpotent.
\item Any configuration of the form $0001^n000$ is stable under $F_1^2$, and
neither $1001$ nor $101$ have antecedents by $F_1$, thus $F_1^t$ at most depends
on the seven center cells.
\item In rule 2, after one step, there can never be two $1$s separated by
less than two $0$s. And, on these configurations, rule 2 is a shift.
\item In rule 4, after one step, the $1$s are all separated by at least
one $0$, and on these configurations, the rule is the identity.
\item Rule 8 is nilpotent.
\item Rule 10 is a left shift on all the configurations with no three consecutive $1$s.
Fortunately, these configurations never appear after one step.
\item For rule 12, the only configurations after one step have only isolated $1$s, on which
this rule is the identity.
\item In rule 19, after two steps, there are no isolated $0$s or $1$s, and
on these configurations, $F_{19}^2$ is the identity.
\item The only difference between rule 24 and the symmetric of rule 2 is
on transition $0 1 1$, which has no antecedent. The same protocol (reverting
the roles of Alice and Bob) can be used, after simulating one step of
the rule.
\item Rule 34 is a left shift on the configurations with no block
of two consecutive $0$s, and these blocks do not have antecedents.
\item For rule 36, we find out by exhaustive search that the only stable pattern
  of length five is $00100$. All other patterns of length five become $0$ after
  two steps.
\item For rule 38, another exhaustive search shows that after one step,
$F_{38}^2$ is equal to the double shift $\sigma^2$.
\item For rule 42, after one step, there are no three consecutive $1$s in
the configuration, and the rule is a left shift on these configurations.
\item Rule 46 is a left shift except on $010$, which has no antecedent,
and $111$, whose antecedents have $010$s. Therefore, after two steps,
this rule is actually a left shift.
\item In rule 72, for any $a$ and $b$, $F_{72}(a0110b)=0110$. But $111$ does
not have antecedents by $F_{72}$, and neither does $010$ by
$F_{72}^2$. Therefore, this rule is the identity after two steps.
\item In rule 76, any block of three cells except $111$ is stable. Therefore,
this block disappears after one step, and the rule becomes the identity.
\item An exhaustive search on all the blocks of length 7 of rule 108 show that
$F_{108}^2$ is the identity on $F_{108}^2(\{0,1\}^{\mathbb{Z}})$.
\item Rule 127 is nilpotent: all cells become $1$ after one step.
\item Rule 138 is a left shift, except on $101$, which has no antecedent
and thus disappears after one iteration.
\item In rule 200, any $0$ is stable (for any $a$ and $b$,
$F_200(a0b)=0$), and so are the blocks of at least two $1$s.
Moreover, isolated $1$s do not have antecedents. Therefore, the rule
depends only on the three central cells.
\end{itemize}
\end{proof}
\end{proposition}
\begin{proposition}
Rule 5 has a protocol for $\textsc{SInv}$ in $O(1)$ bits.
\begin{proof}
For any value of $a$ and $b$, $F_5(a 0 1 0 b)=0 1 0$, and
for any $a,b,c,d$, $F_5^2(a b 0 0 0 c d)\in \{000,010\}$.
Therefore, for the configuration to be invaded, $u$ should
neither contain more than three consecutive $0$s, nor less
than two consecutive $1$s. However, this is not possible after one
iteration of the rule since
$F_5(1 1 0 1 1)=000$, and $F_5(1 1 0 0 1 1)=0 0 0 0$.
\end{proof}
\end{proposition}
\begin{proposition}
Rule 7 has a protocol for $\textsc{SInv}$ in $O(1)$ bits.
\begin{proof}
  First notice that for any values of $w$, $x$, $y$ and $z$, ${F_7^2}(w 1 1 x y
  z) = 1 1$. Since $F_7(0 0 0 0) = 1 1$ and $F_7(0 0 0 1) = 1 1$, a periodic
  word $u$ that would be invaded should have neither blocks of three ore more
  $0$s, nor blocks of two or more $1$s. But since $F_7(0010) = 11$, this leaves
  only one possibility : the word $p$ must be $01$.
  Thus, any perturbation of size $n$ stays at most $n$ bits wide, and
  no invasion can ever occur.
\end{proof}
\end{proposition}
\begin{proposition}
Rule 13 and 29 have a protocol for $\textsc{SInv}$ in $O(1)$ bits.
\begin{proof}
  Let us remark that for any values of $a$ and $b$, $F(a01b) = 01$, for both
  rules. Thus, if the input is different from the periodic background, it can
  only be invaded if the background is equal to $p_1$. But then the last cell
  that is different from the background in the input is a $0$, and this forms a
  wall. Thus, no invasion can occur.
\end{proof}
\end{proposition}
\begin{proposition}
Rule 28 has a protocol for $\textsc{SInv}$ in $O(1)$ bits.
\begin{proof}
First remark that since for any values of $a$ and $b$, $F_{28}(a01b) =
01$. Hence, any periodic background that can be invaded must be
uniform (i.e. have only $0$s or only $1$s). But then the left of the
configuration is necessarily invaded, and the first $01$ or $10$ creates a
wall.
\end{proof}
\end{proposition}
\begin{proposition}
Rule 78 has a protocol for $\textsc{SInv}$ in $O(1)$ bits.
\begin{proof}
  The configurations with no two consecutive $0$s, and with no three
  consecutive $1$s, are stable under this rule. Moreover, $1111$ has no
  antecedent under rule 78, and for all $a$ and $n\geq 2$,
  $F_{78}(a10^n1)=10^{n-1}1$.
  Moreover, in a configuration not containing two consecutive $0$s, blocks
  of exactly three $1$s disappear in one step: indeed, $F(01110)=101$.
  Therefore, if initially, the largest block of $0$s is of length $n$, the
  configuration becomes stable after at most $n+2$ steps (one step to eliminate
  $1111$, $n$ steps for the largest block of $0$, and then one step to
  eliminate $111$).
  Finally, the only pattern that can be invaded is $p_0$, and the presence of a
  $1$ in the configuration is sufficient for it to be invaded, and this can be
  decided with $O(1)$ bits of communication.
\end{proof}
\end{proposition}
\begin{proposition}
Rule 140 has a protocol for $\textsc{SInv}$ in $O(1)$ bits.
\begin{proof}
For any values of $a$ and $b$, $F_{140}(a0b) = 0$. Therefore, the only pattern
that can be invaded contains only $1$s, and it is invaded as soon as the input
contains a $0$, because $F_{140}(110)=0$, and $F_{140}(011)=0$.
\end{proof}
\end{proposition}
\begin{proposition}
Rule 172 has a protocol for $\textsc{SInv}$ in $O(1)$ bits.
\begin{proof}
  For all $a$ and $b$, $F_{178}(a00b)=00$. Therefore, if the background contains
  two consecutive $0$s, it cannot be invaded. Else, remark that for all $a$, $b$
  and $c$, $F_{178}(abc00)=d00$ for some $d$. Therefore, the first $00$ block in
  the input invades the whole configuration. If there is no such block, rule 178
  behaves like a left shift after one step: indeed, the only pattern where it is
  not a left shift is $010$ and patterns containing $00$, but $010$ has no
  antecedents. Therefore, in this case, the configuration is not invaded.
\end{proof}
\end{proposition}
\begin{proposition}
Rule 32 has a protocol for $\textsc{SInv}$ in $O(1)$ bits.
\begin{proof}
  If $p_u$ is different from $p_{01}$, then $F_{32}$ becomes uniformly $0$ after
  $|u|$ steps. Else, if the background pattern is $p_{01}$, and $p_u[x]\neq
  p_u$, then the configuration is invaded with $0$s.
\end{proof}
\end{proposition}
\begin{proposition}
Rule 156 has a protocol for $\textsc{SInv}$ in O(1).
\begin{proof}
  First notice that if the period $u$ is not uniformly $0$ or uniformly $1$,
  then there are walls $01$ around the input $x$, and then $x$ does not invade
  $p_u$. Else, if $u$ is uniform, then $p_u$ is invaded, either to the left if
  $u=1$, or to the right if $u=0$: indeed, the first $0$ (respectively the last $1$)
  of the input forms a wall, and propagates to the left (respectively to the right).
\end{proof}
\end{proposition}
\begin{proposition}
There is a protocol in $O(1)$ for $\textsc{SInv}$ for rule 27.
\begin{proof}
First notice that for all $a$, $b$ and $c$, $F^2( a 1 1 1 b c d ) = 1 1
1$, and $F( a 0 0 0 b ) = 111$.  Thus, if the orbit of $p_u$
contains a block of three $1$s or three $0$s, then no invasion can
occur. Else, an exhaustive exploration of all configurations of size 6
shows that the only possible configurations that do not generate $1 1 1$
or $0 0 0$ are of the form $\{011,001\}\{011,001\}$ (let $A$ be the set
of configurations generated by infinite repetitions of these words).
Moreover, it is easy to notice that this set of configurations is stable under
F, and that for any word $w$ of length 5 of that form, $F^2(w_1 w_2 w_3 w_4 w_5)
= w_5$. Thus, if $p_u[x]$ is still in $A$, then no invasion can occur,
since $F^2$ is a left shift.
\end{proof}
\end{proposition}
\begin{proposition}
Rule 44 has a protocol for $\textsc{SInv}$ in $O(1)$ bits.
\begin{proof}
First remark that for all values of $a$ and $b$, $F_{44}(a 0 0 b)=0 0$. Moreover,
$F_{44}(1 1 1 a)=0 0$ and $F_{44}(0 1 0 a b)=1 1 1$.
Therefore, the only blocks of three letters that do not form walls are $W=\{0 1
1, 1 0 1, 1 1 0\}$. Therefore, the only background pattern that does not form
``walls'' (and thus, that can be invaded) is an infinite repetitions of $011$.
Then, can simply remark that for any change in this pattern introduces a block
of two $0$s. Moreover, for $w\in\{011,101,110\}$, $F_{44}^2(w00)=0$.
Since $0 0$ is a wall, this means that any $x$ such that $p_u(x)\neq p_u$
will invade the configuration.
This condition can be checked with only $O(1)$ bits of communication.
\end{proof}
\end{proposition}
\begin{proposition}
Rules 23, 50, 77, 178 and 232 have a protocol for $\textsc{Pred}$ in
$O(\log n)$ bits.
\begin{proof}
  All these rules create walls on configurations containing either $00$ or $11$,
  or $01$ or $10$. More precisely, for all values of $a$ and $b$:
  \begin{itemize}
  \item $F_{23}(a00b)=11$ and $F_{23}(a11b)=00$.
  \item $F_{50}(a01b)=10$ and $F_{50}(a10b)=01$.
  \item $F_{77}(a01b)=01$ and $F_{77}(a10b)=10$.
  \item $F_{178}(a01b)=10$ and $F_{178}(a10b)=01$.
  \item $F_{232}(a00b)=00$ and $F_{232}(a11b)=11$.
  \end{itemize}
  The proof is the same for all the cases; we do it for rule 23: Alice can send
  the position of her first $00$ or $11$, and one bit indicating whether it is a
  $00$ or a $11$. Bob then knows the only relevant part of her configuration (an
  alternation of $0$s and $1$s), and can compute the result.
  This protocol requires $O(\log n)$ bits of communication.
\end{proof}
\end{proposition}
\begin{proposition}
Rules 40, 130, 162 and 168 have a protocol for $\textsc{Pred}$
in $O(1)$ bits.
\begin{proof}
First notice that in all four rules, for all $a,b\in\{0,1\}$, $f(ab0)=0$.
Therefore, if Bob has one $0$, he can predict the result alone. Else,
only one bit is needed to inform Alice that he has only $1$s.
\end{proof}
\end{proposition}
\begin{proposition}
There is a protocol in $O(1)$ for $\textsc{SInv}$ for rule 104.
\begin{proof}
  Let us first notice that rule 104 is symmetric, and that for all $a$ and $b$,
  $F_{104}(a 0 0 b) = 0 0$. Moreover, $F_{104}(1111)=00$.  Therefore, any
  configuration without walls only contains blocks of one or three $1$s, or
  $0$s alone.
  Now, $F_{104}(010111)=0110$, and $F_{104}(0111010)=10110)$. Both configurations
  contain $0110$. However, if $0110$ appears, it is necessarily surrounded by $1$s,
  and $F_{104}^2(101101)=00$.
  Therefore, the only $p_u$ without walls are $p_{01}$ and $p_{0111}$. Therefore,
  any change in the configuration creates a wall, after which Alice or Bob can decide
  whether their part of the configuration is invaded, and communicate this information
  using $O(1)$ bits.
\end{proof}
\end{proposition}
\begin{proposition}
There is a protocol in $O(\log n)$ for $\textsc{Pred}$ of rule 132.
\begin{proof}
For any $a,b \in \{0,1\}$, $F_{132}(a 0 b)=0$. Thus, Alice only needs to send
the length of the longest string of $1$s she has from the center, and Bob
can compute the relevant bits of the configuration.
\end{proof}
\end{proposition}
\begin{proposition}
There is a protocol in $O(1)$ for $\textsc{SInv}$ for rule 152.
\begin{proof}
There are
two cases:
\begin{itemize}
\item Either $p_u=p_1$, in which case it is invaded to the left, since the
  rightmost $0$ of $x$ creates a vertical wall (because $F_{152}(a011)=01$, and
  $F_{152}(111)=1$), and its leftmost $0$ propagates to the left.
\item Else, let $n$ be the size of the largest block of $1$s in $p_u[x]$.  For
  all $a$, $F_{152}(a011)=01$, and $F_{152}(110)=0$. Therefore, $F_{152}^n(p_u[x])$
  does not contain the pattern $11$.
  Thus, a simple observation of the rule shows that on these configurations, it
  is a shift to the right, and therefore the configuration cannot be invaded.
\end{itemize}
Detecting the case takes $O(1)$ bits of communication.
\end{proof}
\end{proposition}
\begin{proposition}
Rule 156 has a protocol in $O(1)$ for $\textsc{SInv}$.
\begin{proof}
  First notice that $0 1$ is a wall in rule 156: for all $a,b\in\{0,1\}$,
  $F_{156}(a 0 1 b)=0 1$. Thus, the only case where invasion could occur are
  when the background pattern has only $0$s or only $1$s (else, a wall
  appears on both sides). If there are only $0$s, the first $1$ creates a wall,
  and since $f_{156}(1 0 0)=1$, the right of the configuration get invaded by
  the last $1$. Since $f(1 1 0)=0$, the same happens when the background pattern
  has only $1$s.
\end{proof}
\end{proposition}
\begin{proposition}
Rule 184 has a protocol in $O(\log n)$ for $\textsc{Pred}$, and
this protocol is optimal.
\begin{proof}
  Consider the blocks of two cells in rule 184. Let $A=0 0$, $B=0 1$, $C=1 0$
  and $D=1 1$. Then, for all $n\geq 0$:
  $$F_{184}^n(A\{B,C\}^n)=A$$
  $$F_{184}^n(\{B,C\}^n D)=D$$
  $$F_{184}^n(A D)=B$$
  $$F_{184}^n(D A)=B$$
  Intuitively, this means that the dynamics of this rule has two particles, one
  moving towards the right, the other towards the left, and any collision
  destroys them.
  Thus, let $|A_{Alice}|$ be the number of $A$ Alice has, $|D_{Alice}|$ her
  number of $D$s, $|A_{Bob}|$ the number of $A$ Bob has, and $|D_{Bob}|$ his
  number of $D$s.
  Moreover, we say that position $i$ is \emph{free} if:
  $$|D(w_0…w_i)| \geq |A(w_0…w_i)|$$
  $$|D(w_{i+1}…w_n)| \geq |A(w_{i+1}…w_n)|$$
  Then, the following protocol solves $\textsc{Pred}$ for rule 184:
  \begin{itemize}
  \item Alice sends $N_A=max(0,|A_{Alice}| - |D_{Alice}|)$ to Bob.
  \item If $N_B>N_A$, then Bob knows the answer (if he has a $C$ particle in
    a free zone, the result is $C$, else it is $B$).
  \item Else, if $N_B<N_A$, then Alice knows the answer: if she has a $C$
    particle in a free zone, then the result is $C$, else it is $B$.
  \end{itemize}
  The following fooling set shows that this protocol is optimal:
  $$S=\{A^i B^{n-i}, B^{n-i}D^i \| i\in \{1,\ldots,n\}\}$$
\end{proof}
\end{proposition}
With a slight modification, the protocol we had for rule 184 can also
predict rule 56:
\begin{proposition}
There is a protocol in $O(1)$ for $\textsc{Pred}_{F_{56}}$.
\begin{proof}
  Using the same rescaling, there are only two differences:
$$F(DD)=A\hbox{ and }F(BD)=C$$
But none of these two ``problems'' have any antecedent, thus they disappear
after one step. Only $O(1)$ bits of communication are needed to simulate this
step.
\end{proof}
\end{proposition}
\subsection{The last candidates to universality}
In the last section, we have shown simple protocols for a large number of
elementary cellular automata, and essentially problems $\textsc{Pred}$ and
$\textsc{SInv}$. The proof for rule 94 is more complex, and appears in
\cite{goles2011}.
The status of the following 33 cellular automata remains open: 3, 6,
9, 11, 14, 18, 22, 25, 26, 30, 33, 35, 37, 41, 43, 45, 54, 57, 58, 62,
73, 74, 106, 110, 122, 126, 134, 142, 146, 152, 154, 164, 204.
\section{Perspectives}
This work opens new perspectives on the analysis of natural systems: indeed,
this is the first systematic proof that a large class of cellular automata, not
chosen on purpose, is simple. Open questions include the proof of lower bounds
on the remaining systems: is there a simple method to prove them? an algorithmic
one?
Moreover, it might become possible at some point to apply this method to other
theoretical, abstract systems. However, a real challenge opened by our results
is the applicability of these techniques to real-world data, in particular from
biological systems, for instance neurons or the evolution.
\bibliographystyle{plain}

\end{document}